\documentclass[twocolumn,twoside,slac_two]{revtex4}
\usepackage{graphicx}
\usepackage{fancyhdr}
\usepackage{amsmath}
\usepackage{aas_macros}

\begin{document}

\title{Extreme TeV Blazars and Lower Limits on Intergalactic Magnetic Fields}

%

\author{Timothy C. Arlen} 
\affiliation{Department of Physics and
  Astronomy, University of California, Los Angeles, CA 90095, USA }

\author{Vladimir V. Vassiliev} 
\affiliation{Department of Physics and
  Astronomy, University of California, Los Angeles, CA 90095, USA }

\begin{abstract}
The intergalactic magnetic field (IGMF) in cosmic voids can be
indirectly probed through its effect on electromagnetic cascades
initiated by a source of TeV gamma rays, such as blazars, a subclass
of active galactic nuclei. Blazars that are sufficiently luminous at
TeV energies, ``extreme TeV blazars'', can produce detectable levels
of secondary radiation from inverse Compton scattering of the
electrons in the cascade, provided that the IGMF is not too large. We
reveiw recent work in the literature which utilizes this idea to
derive constraints on the IGMF for three TeV-detected blazars-1ES
0229+200, 1ES 1218+304, and RGB J0710+591, and we also investigate
four other hard-spectrum TeV blazars in the same framework. Through a
recently developed detailed 3D particle tracking Monte Carlo
simulation code, incorporating all major effects of QED and
cosmological expansion, we research effects of major uncertainties
such as the spectral properties of the source, uncertainty in the
intensity of the UV - far IR extragalactic background light (EBL),
under-sampled Very High Energy (VHE; energy $\geq$ 100 GeV) coverage,
past history of gamma-ray emission, source vs. observer geometry, and
jet AGN Doppler factor. The implications of these effects on the
recently reported lower limits of the IGMF are thoroughly examined to
conclude that presently available data are compatible with a zero IGMF
hypothesis.
\end{abstract}

\maketitle

\thispagestyle{fancy}


\section{Introduction}\label{section_intro}
It is well known that astrophysical magnetic fields are ubiquitous in
galaxies and galaxy clusters on the order of a micro Gauss (see
e.g.~\cite{GrassoRubinstein01, Widrow02}). Furthermore, there is
increasing theoretical evidence based on numerical simulations of
structure formation that nano Gauss order fields permeate filaments of
the large scale structure (see e.g.~\cite{Ryu_2008}). However, at
present there has been no detection of the intergalactic magnetic
field (IGMF) presumed to exist in the Cosmic Void regions of the large
scale structure. The detection of the IGMF could provide important
insights for solving outstanding problems of its origin and role in
both the cosmology and astrophysics of structure formation
\citep{GrassoRubinstein01,Zweibel2006,Widrow02,Kronberg94}.

Until recently, only upper limits on $B_{IGMF}$ have been established,
on the order of $B_{IGMF} \sim 10^{-9}$ Gauss, at a correletion length
of 1 Mpc, and this limit weakens as the correlation length
($\lambda_c$) decreases as $\sim$ $\lambda_c^{-1/2}$
\citep{NeronovSemikoz_Sensitivity09}. However, a more sensitive
technique has emerged during the past few years which may become a
tool for the measurement of IGMF characteristics. This technique
relies on observations of blazars, in the energy range from 100 MeV to
greater than 10 TeV and is described by several authors
\citep{NeronovSemikoz06, Eungwan09, Dolag09, ENS09}. Briefly, TeV-scale
gamma rays from the blazar interact with the UV to far-IR
extragalactic background light (EBL), producing electron-positron
pairs, which then undergo inverse Compton (IC) scattering on the CMB
photons, producing secondary gamma rays of a lower energy than the
primary. Because the pairs' trajectories depend on the interactions
with the magnetic field, comparing the GeV-scale secondary gamma ray
radiation with existing data can be used to characterize properties of
the IGMF.

In the past few years, a number of studies \citep{Neronov2010,
  Tavecchio2010, Dermer2010, HaoHuan2011, Dolag2010, Taylor2011} have
demonstrated that a lower limit on the IGMF strength can in principle
be derived by requiring that the secondary gamma ray GeV radiation
which would be produced at a given $B_{IGMF}$ not exceed the measured
value. A consensus has begun to emerge that a minimum value of
$B_{IGMF} \approx 10^{-17} - 10^{-18}$ Gauss is required to explain the
observations. With a newly developed 3-dimensional Monte Carlo code
combined with a detailed and general statistical analysis method, the
purpose of these proceedings is to show that the current observational
data is consistent with the $B_{IGMF}$ = $0$ hypothesis, when the
systematic and statistical uncertainties are accounted for.

\section{Monte Carlo Simulations}\label{section_sims}
In order to explore in detail the potentially observable effects of
cascading in the voids of the large scale structure, we have developed
a fully 3-dimensional Monte Carlo code, which propagates individual
particles of the cascade in a cosmologically expanding Universe and
accounts for the QED interactions with the EBL and CMB without
simplifications.

The spectral energy density of the EBL from the UV to the far-IR is
not precisely known, and therefore we choose to model it with specific
SEDs close to the energy density of the recent EBL model of
\cite{Dominguez_ebl_2010}.  The magnetic field is described as a
system of cubic cells with an edge length set to 1 Mpc, with equal
field amplitudes in each cell, but randomly oriented in direction.

The gamma ray source model employed is based on leading theoretical
speculations about the nature of the TeV blazar source (see
e.g. \cite{Urry_Padovani_1995}). In the reference frame of the blazar
jet, GeV - TeV photons are distributed isotropically with a broken
power law spectrum of index $\alpha$ for the higher energy component
and $\gamma$ for the lower energy component. Once this distribution is
boosted into the reference frame of the host galaxy with a doppler
boost factor of $\Gamma = \left(1 - \beta^2 \right)^{-1/2}$, the
observed differential flux energy density (dFED) can be parameterized
as

\begin{equation}\label{broken_power_law_model}
  \epsilon\frac{dF}{d\epsilon} = F_0\delta^{2}
\begin{cases}
\left( \frac{\epsilon}{\epsilon_B\delta}  \right)^{-\gamma+1} 
  \text{exp}\left(\frac{-\epsilon}{\epsilon_{c}} \right)  & 
  \frac{\epsilon}{\epsilon_B\delta} < 1 \\
  \left( \frac{\epsilon}{\epsilon_B\delta}
  \right)^{-\alpha+1}
  \text{exp}\left(\frac{-\epsilon}{\epsilon_{c}} \right) &
  \frac{\epsilon}{\epsilon_B\delta} > 1
\end{cases}.
\end{equation}

\noindent where $\epsilon$ is the photon energy in the reference frame
of the host galaxy, $\delta$ = $\left[\Gamma \left(1 -
  \beta\cos\theta_{\text{v}} \right)\right]^{-1}$, $\theta_{\text{v}}$
is the viewing angle from the blazar jet axis to the line of sight of
the observer, $F_o$ is a flux normalization factor, $\epsilon_c$ is
the exponential cutoff energy, and $\epsilon_B$ is the spectral break
energy. This six parameter $\gamma$-ray source spectrum is given at
the redshift of the host galaxy and is necessary and sufficient to
satisfy observational data of TeV blazars in both the HE
(\emph{Fermi}-LAT) and VHE (IACT) regimes, and it is more general than
previous studies which only considered single intrinsic power law
models for the primary gamma ray emission, over the energy range
spanning more than five orders of magnitude from $\approx 100$ MeV to
$> 10$ TeV.

\section{Testing the $B_{IGMF} = 0$ hypothesis}
\begin{figure*}[t!]
  \includegraphics[scale=0.15]{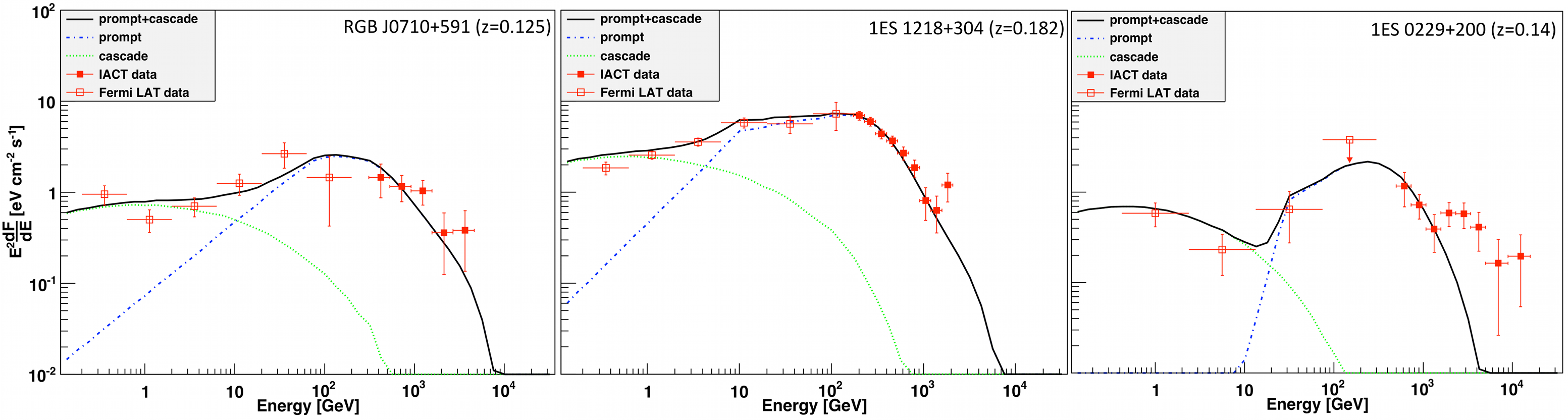}
  \caption{\label{best_fit_model} Best fit models for (a) (left)
    source RGB J0710+591 where $\alpha$ = 1.8, $\epsilon_C$ = 3.16
    TeV, which is inconsistent with the $B_{IGMF} = 0$ hypothesis at
    $<$ 90\% (b) (middle) source 1ES 1218+304 where $\alpha$ = 1.8,
    and $\epsilon_C$ = 3.16 which is inconsistent with the $B_{IGMF} =
    0$ hypothesis at $<$ 80\% (c) (right) source 1ES 0229+200 where
    $\alpha$ = 1.3, $\epsilon_C$ = 1 TeV, which is inconsistent with
    the $B_{IGMF} = 0$ hypothesis at $>$ 99\%. }
\end{figure*}

The most comprehensive work yet on constraining the IGMF was done by
\cite{Taylor2011} (hereafter referred to as TVN11), in which three
sources were found (RGB J0710+591, 1ES 1218+304, and 1ES 0229+200)
whose combined quasi-simultaneous IACT and Fermi-LAT dFED are not
compatible with the $B_{IGMF} = 0$ hypothesis at greater than $95 \%$
confidence level. It was suggested that $B_{IGMF} \geq 10^{-17}$ G is
required to explain observations.

With the newly developed simulation and analysis code, we re-examined
the conclusion of a lower limit on $B_{IGMF}$. The spectral model of
Eq \ref{broken_power_law_model} was fit to the same IACT data as used
in TVN11, but with about one more year of \emph{Fermi} data. A large
range of parameter space of the model was scanned over ($\alpha$,
$\epsilon_c$, $F_o$, $\gamma$, $\epsilon_B$), while several parameters
were left fixed, including 1) a fixed source-observer geometry of
$\Gamma = 10$ and $\theta_v = 0^{\circ}$, 2) the EBL model with an SED
very close to \cite{Dominguez_ebl_2010}, 3) the duty cycle of the TeV
emission, and 4) assuming no structures with an enhanced magnitude of
magnetic field (for ex: galaxy clusters, filaments, etc.)  between the
source and observer. For each model, a $\chi^2$ fit to the combined
IACT/\emph{Fermi} data was performed, and also the corresponding
confidence level for rejecting the $B_{IGMF} = 0$ hypothesis. (For
more details on this procedure, as well as the chosen data sets,
please see \cite{Arlen_2012}). The best fitting models for each source
are shown in Fig~\ref{best_fit_model}. The first two sources, RGB
J0710+591 and 1ES 1218+304 (Fig~\ref{best_fit_model} a and b) were
found to be incompatible with the $B_{IGMF} = 0$ hypothesis at $<$ 90
\%, which does \emph{not corroborate the conclusion of TVN11 that the
  $B_{IGMF} = 0$ hypothesis is rejected for these sources}. It was
found that a combination three factors were the primary reasons for
this discrepancy, which are the following: 1) the updated Pass 7
\emph{Fermi}-LAT data used in the present work, 2) the more general
description of the broken-power law source model in the analysis of
the data, and 3) the more robust statistical analysis which used
information about the cascade flux to define the spectral index in
each bin as an input to the Fermi likelihood spectral analysis.

\begin{figure*}[t!]
  \includegraphics[scale=0.46]{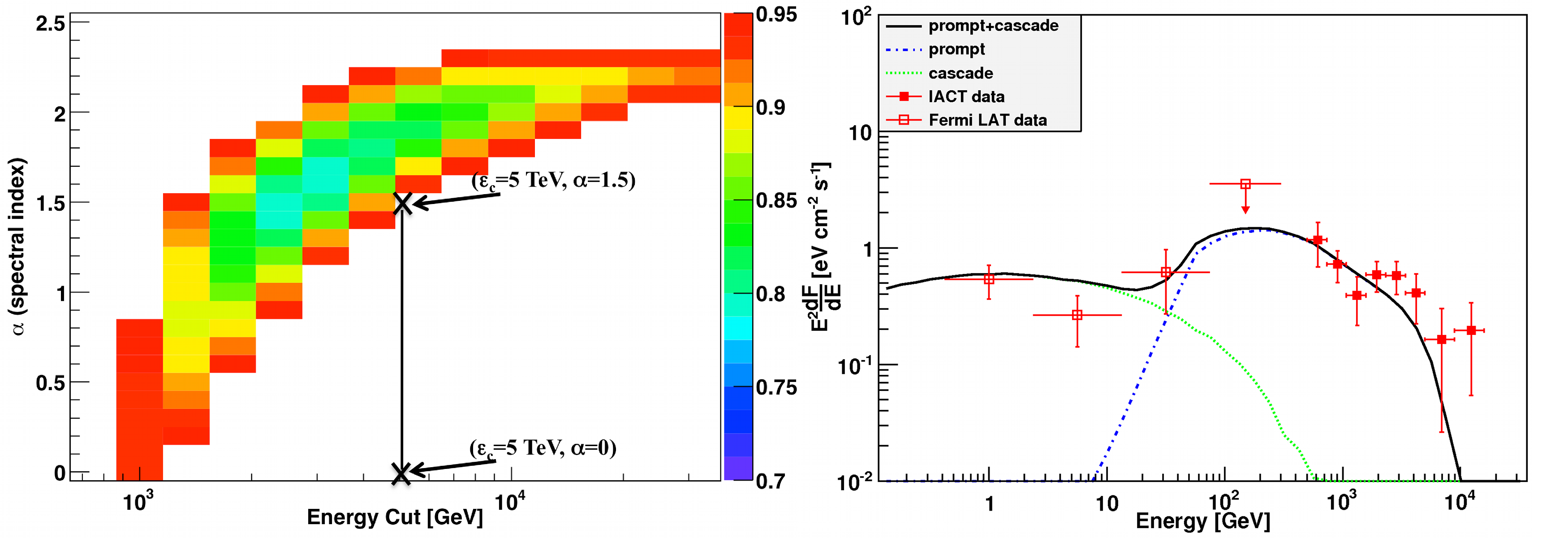}
  \caption{\label{1es0229_figure} (a) (left) Confidence level for the
    simulated source models for 1ES 0229+200 obtained under the
    assumption of a low EBL Model close to the lower limits due to
    galaxy counts. The black line terminated with crosses represents
    the range of models analyzed in \cite{Vovk_EBL_IGMF_2012}. (b)
    (right) dFED of 1ES 0229+200 for the best fitting model at $\alpha
    = 1.6$, $\epsilon_c = 3.16$ TeV.}
\end{figure*}

It is confirmed that the data for 1ES 0229+200 are incompatible with
the $B_{IGMF} = 0$ hypothesis at $> 99 \%$ confidence level, with the
standard assumptions of 1) - 4) (listed in the previous
paragraph). Since only this single source was found to be in conflict
with a non-zero IGMF, each of these standard assumptions were varied
to within their uncertainty limits to research all the possible ways
in which the non-zero IGMF requirement can fail. For example, the duty
cycle assumption was varied, because the 3 year time-averaged
\emph{Fermi}-LAT spectrum is combined in the spectral model fits with
the VHE spectrum which is averaged over a much shorter time
period. Thus, it is necessary to make an assumption regarding the
stability of the pointed VHE observations (duty cycle).  To
investigate the effect of a reduced duty cycle, the spectrum of this
source was modified at the highest 5 energy points to half an order of
magnitude of their reported values. It was found that the VHE--HE data
set combined in this way does not rule out the $B_{IGMF} = 0$
hypothesis to more than 95 \% confidence level. Therefore, if the true
3-year time-averaged spectrum is over-estimated by about 50 \%, then
compatibility with the $B_{IGMF}$ = $0$ hypothesis would be
achieved. A similar compatibility could be achieved if the location of
the source within a filament is separated by at least a few hundred
Mpc from the nearest Cosmic Void. The enhanced magnetic field within a
filament could potentially isotropize the electrons before IC
scattering occurs along the line of sight to the obsverver and could
also explain the observations without invoking a non-zero IGMF.

Alternatively, the energy density of the EBL in the near IR band can
be reduced to the lower limits derived from galaxy counts data to
avoid incompatibility with the zero IGMF assumption. This latter case
of EBL uncertainty has also been investigated recently
\citep{Vovk_EBL_IGMF_2012} to conclude that this data provides joint
constraints on the IGMF and EBL in which the zero field is in the
exclusion region. Figure~\ref{1es0229_figure} illustrates our finding
that this conclusion is the result of a limited exploration of the
parameter space of uncertainties in the intrinsic properties of the
source. Therefore, for EBL models which are very low in the near-IR
energy density but still allowable by the currently established lower
limits, the data of 1ES 0229+200 can be reconciled with a zero
magnetic field ($B_{IGMF} = 0$).

\section{Discussion}
We have investigated the HE--VHE energy spectrum of several extreme
TeV blazars for which radiation in the HE band may be dominated by the
secondary photons produced through cascading in the voids along the
line of sight. These sources are characterized by their observed hard
spectra in the VHE band accompanied by redshifts of order $\approx$
0.1 suggesting a very large energy output into pair production and
subsequent cascading.  For these sources, observations in the HE band
can therefore limit the flux of secondary photons and establish a
lower bound on the IGMF.  This strategy has been utilized in several
publications \citep{Neronov2010, Taylor2011, Tavecchio2010, Dermer2010,
  Dolag2010, HaoHuan2011} to suggest $B_{IGMF} \leq 10^{-17} -
10^{-18}$ G in the local cascading environment. In contrast to these
studies, we systematically investigated effects of a wide range of
uncertainties using detailed 3D Monte Carlo simulations to conclude
that the $B_{IGMF} = 0$ hypothesis remains compatible with current
observations.

\bigskip 
\begin{acknowledgments}
  This research is supported by a grant from the U.S. National Science
  Foundation (PHY-0969948). All simulations were performed on the UCLA
  hoffman2 cluster.
\end{acknowledgments}

\bigskip 



\end{document}